\documentclass[sn-mathphys-num, iicol]{sn-jnl}

\usepackage{graphicx}%
\usepackage{multirow}%
\usepackage{amsmath,amssymb,amsfonts}%
\usepackage{amsthm}%
\usepackage{mathrsfs}%
\usepackage[title]{appendix}%
\usepackage{xcolor}%
\usepackage{textcomp}%
\usepackage{manyfoot}%
\usepackage{booktabs}%
\usepackage{algorithm}%
\usepackage{algorithmicx}%
\usepackage{algpseudocode}%
\usepackage{listings}%
\usepackage[utf8]{inputenc}

\theoremstyle{thmstyleone}%
\theoremstyle{thmstyletwo}%

\theoremstyle{thmstylethree}%

\begin{document}

\title[Article Title]{Connectivity Aware and Energy Efficient Self-Organizing Distributed IoT Topology Control}


\author[1]{\fnm{Azra} \sur{Seyyedi}}\email{azra.seyyedi@iasbs.ac.ir}

\author[1]{\fnm{Sina} \sur{Dortaj}}\email{dortaj.sina@gmail.com}

\author[2, 3, 4]{\fnm{Mahdi} \sur{Bohlouli}}\email{me@bohlouli.com}

\author*[1, 3]{\fnm{SeyedEhsan} \sur{Nedaaee Oskoee}}\email{nedaaee@iasbs.ac.ir}

\affil*[1]{\orgdiv{Department of Physics}, \orgname{Institute for Advanced Studies in Basic Sciences (IASBS)}, \orgaddress{\street{Prof Yousef Sobouti Blvd}, \city{Zanjan}, \postcode{45137-66731}, \state{Zanjan}, \country{Iran}}}

\affil[2]{\orgdiv{Department of Computer Science and Information technology}, \orgname{Institute for Advanced Studies in Basic Sciences (IASBS)}, \orgaddress{\street{Prof Yousef Sobouti Blvd}, \city{Zanjan}, \postcode{45137-66731}, \state{Zanjan}, \country{Iran}}}

\affil[3]{\orgdiv{Research Center for Basic Sciences and Modern Technologies (RBST)}, \orgaddress{\street{Prof Yousef Sobouti Blvd}, \city{Zanjan}, \postcode{45137-66731}, \state{Zanjan}, \country{Iran}}}

\affil[4]{\orgdiv{Research and Innovation Department}, \orgname{Petanux GmbH}, \orgaddress{\street{Dottendorfer}, \city{Bonn}, \postcode{53129}, \state{North Rhine-Westphalia}, \country{Germany}}}


\abstract{
Internet of Things with distributed structure has pervaded every area of modern life. 
One of the important applications that they can have is wireless communication under emergency circumstances.
In order for a network to be functional in this scenario, it must be developed without the aid of a pre-established structure and operated in a self-organized manner to accommodate the communication requirements of the time. Although the design and development of such networks can be highly advantageous, they frequently confront difficulties, the most significant of which is attaining and maintaining effective connectivity to have reliable communications despite the requirement to optimize energy usage. In this study, we present a model for self-organizing topology control for ad hoc-based internet of things networks that can address the aforementioned challenges. The model that will be presented employs the notion of the Hamiltonian function in classical mechanics and has two key objectives: regulating the network's topology and dynamics to enhance connectivity to a desirable level while requiring the least amount of energy possible. The results of the extensive simulations indicate that the proposed model satisfactorily fulfills the goals of the problem. }




\maketitle

\section{Introduction}\label{sec:introduction}

Distributed IoT networks are the purest form of IoT.
Together, the constituent members in a distributed network share resources among themselves and form a interconnected network in such a way that each of them has integrity and accessibility to existing data.
By decentralizing from a central authority, the formulation of a distributed approach enables the running of software in parallel across devices and close to where calculations are actually required. As a result, data transfer between members is minimized to the greatest extent possible, and the new approach allows for the implementation of more devices. In comparison to the centralized strategy, this approach leads to a more stable network with greater fault tolerance and possibly that is more time-optimal. 
In short, IoT technology, which is used in people's daily lives, needs to be transparent, secure, dependable, and powerful. It appears that these features can be added to the IoT network via the distributed structure.

A suitable network environment or a proper backbone for distributed IoT can be provided by ad hoc networks. In fact, IoT devices can be designed to form ad hoc networks when needed.
Devices in such architectures communicate with each other on-demand and directly in a peer-to-peer (P2P) manner. 
In certain scenarios, where there are mobile IoT devices in the network, they can make use of a Mobile Ad hoc NETwork (MANET) to establish communications.
MANETs share self-organizing feature \cite{srilakshmi2021improved, khan2020game, sharma2021comprehensive} and it is due to the independent mobility of each device, along side the fundamental characteristics of the ad hoc network including open environment, distributed collaboration, dynamic topology, and limited capability \cite{deng2002routing}. 

Absence of a reliable communication foundation is a crucial challenge when implementing a practical MANET. This desired reliability and efficiency can be impacted by a variety of factors. The more devices that can be connected, and the more stable these connections are, the more efficient the communications on MANET will be. The network connectivity may be approximated and used to explain the first instance, while the network lifetime \cite{lata2020fuzzy, esmaeilpour2018connectivity} and the caliber of each connection can be used to support the second one. 

A network's connectivity is established by the dissemination of information over it. 
Each distinct IoT member contributes to the connectivity of the whole network. 
The probability of communication between devices and consequently the connectivity of the network grow as transmission power of each member increases. However, merely boosting each device's transmission power is not the solution to the routing and wireless connectivity issues in MANETs. 
Given that the majority of sensors use batteries as their energy source, which has a finite capacity, an increase in devices' energy consumption shortens the network's lifetime \cite{cardei2005improving}. 
High power also causes excessive interference and a breakdown in communications \cite{von2014transmission}. So limiting potential interference and power usage is just as crucial as managing network connectivity. Reliability and efficiency can conflict when there is a limited supply of energy \cite{lin2016atpc}.

In traditional networks, devices transmit at maximum power, and routing protocols implicitly construct the architecture. Contrary to this, IoT devices in a wireless network can collectively decide their transmission power and control the IoT topology by building a suitable neighborhood under some circumstances \cite{li2005design}. Given how network topology affects its performance, it would seem that improving IoT network performance and addressing the aforementioned issues can be accomplished by consciously managing network topology. 

The idea if topology control is to alter the network topology dynamically in different possible ways, while maintaining certain network characteristics like connectivity, energy cost, and interference \cite{chiwewe2011distributed}. Therefore, topology management can be viewed as a trade-off between global connectivity of network in one hand, and energy usage minimization, as well as interference decrease on the other \cite{burkhart2004does}. That means energy requirement and signal interference can be diminished in the best way possible by selecting the lowest transmitting power contingent on preserving network connectivity \cite{li2005design}.

In this study, we propose a novel self-organizing method for topology control of ad hoc-based IoT networks, which can address some of the discussed drawbacks and difficulties. With this method, the goals of the IoT topology control are concentrated on two major problems: to maximize network connectivity while also minimizing energy usage. These two variables are simultaneously optimized until they reach a relative balance and the network reaches its stable state.  
By achieving the goal of increasing network connectivity, the amount of time needed to transfer data and the amount of data lost during transfer are decreased, and the network's reliability is raised indirectly. Additionally, the network's lifetime is extended by meeting the goal of minimizing energy usage. The issue of radio interference will ultimately be indirectly lessened to some degree with the achievement of these goals.

A physics-based approach serves as the foundation for topology control mechanism presented for distributed IoT in this paper. Actually, this novel method is put forth by utilizing the Hamiltonian mechanics in classical physics and taking into consideration MANETs' inherent self-organization property. The network under study is regarded as a complex physics system, and a Hamiltonian function is established for that system by taking into account the system's structural characteristics and existing connections. Hamiltonian function describes how conservative physical systems evolve {\color{black} and attain steady state} \cite{shankar2012principles, kozlov1983integrability, bertalan2019learning}.

{\color{black}In this study, Hamiltonian function is used as a cost function and its value is minimized by a numerical method.}
The formulation of the Hamiltonian function and the choice of its coefficients are done in a way that will cause the dynamics of the network to proceed in a manner that will transform the topology of the network (apart from its initial form) into the desired topology. The final topology must, in addition to satisfying the two main objectives listed above, also be coincident with the physical steady state of the system, which is what is meant by the desired topology in this context.

The proposed method of this article is applied to internet of things networks with centralized, distributed ad hoc, and distributed mobile ad hoc structures in 2 and 3 dimensions. This method is able to transform a network?s topology from a fully discrete configuration to one that is almost connected(connectivity $\geq 95\% $). This connectivity is robust to small-scale local changes in the network and also it can quickly restore the network's strong connectivity against larger-scale events like being attacked or failure of a significant percentage of all nodes. 
This method, while ensuring the network connectivity, decreases energy consumption by adjusting the nodes' transmission range to the proper value. 
To the best of our knowledge, this work is the first network topology control approach that makes use of the Hamiltonian function notion. 

In our ad hoc networks, both the network structure and the proposed control mechanism are completely distributed. Every node has an identical role within the network and functions independently. This feature provides the basis for self-organization of distributed network behavior. The Hamiltonian method allows the network's devices to cooperate and function together while calculating and using the least amount of data required to accomplish the network's objectives.
There isn't any specific limitations, such as those pertaining to each node's degree, the number of links it has, its transmission range, etc.
It is a general, flexible, and scalable model that can be adapted to different environments and expanded for a range of 2D and 3D applications. It can be employed with a variety of routing protocols.
Extensive simulations have validated these statements.

The rest of this paper is structured as follows. Section \ref{sec:relared_work} provides an overview of the related researches. Section \ref{sec:methodology} gives a comprehensive methodology of the proposed model. Results of the simulation analysis are reported and discussed in section  \ref{sec:results}. Finally, in section \ref{sec:conclusion} the conclusions of the paper are described, along with a few ideas for future studies.

\section{Related Work} 
\label{sec:relared_work}
The topology of an IoT network can be altered dynamically in a variety of ways, while maintaining certain network characteristics like connectivity, low energy cost, and etc. There have already been numerous approaches presented for controlling the topology of IoT networks. We concentrated on wireless ad hoc and sensor IoT networks in this paper. The state-of-the-art solutions that have been presented to tackle some issues with topology control in wireless ad hoc and sensor networks are briefly reviewed in this section. 

One key method that many algorithms use to control the topology of the network is power control. In such a method, each device is given sufficient transmission power to accomplish a network-wide property. As a consequence, links for communication will be established or eliminated in the network.
Another important category of methods is focused on management of wireless interfaces. In this category, the wireless interface is managed along with the active, sleep, and powered off modes of each device to carry out the topology arrangement. This alters the topology between devices with time. 

There is also other popular category of significant algorithms based on mobility. This type of topology control involves the intended relocation of some devices in order to establish different interconnections. Table \ref{tab:related_work} provides a summary of several studies carried out using one of these three categories of methodologies. The various research studies done in this area may combine the described methodologies or maybe take an entirely different approach. According to several traits, different other categories might be proposed for the topology control algorithms.

\begin{table}[h]
	\caption{An overview of some related research studies.}\label{tab:related_work}%
	\footnotesize
	\begin{tabular}{ |p{1.7cm}||p{5cm}|  } 
		\toprule
		category of methods & publications  \\
		\midrule
		power management    & \cite{burkhart2004does} \cite{chaudhry2018optimized} \cite{devika2020power} \cite{moaveninejad2005low} \cite{krunz2004transmission} \cite{nasir2016distributed} \cite{genda2020topology} \cite{ ramanathan2000topology} \cite{gomez2006variable}  \cite{jornet2008distributed} 
		\cite{quiroz2010energy} \cite{park2003adaptive} \cite{xu2012energy} \cite{li2001minimum} \cite{kim2014dynamic} \cite{liu2002mobilegrid} \cite{wattenhofer2001distributed} \cite{bahramgiri2006fault} \cite{huang2002topology} \cite{blough2003k} \cite{wang2003topology} \cite{lloyd2002algorithmic} \cite{sathyamanagalam2011topology} \cite{liu2003distributed} \cite{li2001analysis} \cite{montemanni2005swarm} \cite{kim2009improving} \cite{blough2002symmetric} \cite{carvalho2010cross} \cite{pradhan2010adaptive} \cite{yadav2011novel} \cite{borbash2002distributed} \cite{femila2014transmission}    \\
		\hline
		interface management    & \cite{schurgers2002stem} \cite{ye2003peas} 
		\cite{cerpa2004ascent} \cite{deng2005scheduling} \cite{schurgers2002topology} \cite{li2015low} \cite{harris2009idle} \cite{xing2005integrated}  \cite{kumar2004k} \cite{sanchez2011low} \cite{ wills2006low} \cite{ chen2001span} \cite{raja2016secured} \cite{singh1998pamas} \cite{ su2016energy} \cite{zorzi2010effects} \cite{coutinho2015modeling} \cite{coutinho2016modeling} \cite{zhang2005maintaining} \cite{polastre2004versatile} \cite{tian2002coverage} \cite{zheng2003asynchronous} \cite{jiang2004coverage} \cite{tseng2003power} \cite{buettner2006x} \cite{liu2009cmac} \cite{xu2001geography} \cite{xu2003topology} \cite{ding2008adaptive} \cite{zhou2007sleep} \cite{yi2016energy} \cite{desikan2020topology} \cite{feeney2001investigating}   \\
		\hline
		mobility management    & \cite{coutinho2013dcr} \cite{coutinho2015novel} \cite{basha2014analysis} \cite{gjanci2017path} \cite{basha2014analysis} \cite{o2012multi} \cite{forero2014rollout} \cite{khan2014greedy} \cite{khan2015scheduling} \cite{khan2016optimizing} \cite{jaffe2006sensor} \cite{wang2006movement} \cite{ganeriwal2004self} \cite{wang2007bidding}  \cite{li2003communication} \cite{bui2001randomized} \cite{ou2004connecting} \cite{chandrashekar2004providing} \cite{han2006smart} \cite{hauert2008ant}    \\
		\botrule
	\end{tabular}
\end{table}

As our suggested approach falls under the first category, the remainder of this section will include a discussion of some of the works that corresponds to this area. The authors in \cite{burkhart2004does} offer a straightforward clarification of interference. They provide spanner and connectivity-preserving interference-minimal designs based on this notion. They introduce a distributed local algorithm (LLISE) that determines an interference-optimal spanner structure in order to satisfy the criterion that the final topology should be a spanner with a particular stretch factor. Optimized Power Control (OPC), a clustering-based topology control technique, is presented in \cite{chaudhry2018optimized} which employs the modified Gabriel Graph (GG) algorithm to calculate the optimum transmission power of nodes in order to maintain network connectivity. This work uses the weight of nodes for clustering. The node degree, existing battery power, and relative mobility are used to calculate each node's weight in this article.

Article \cite{devika2020power} presents a hybrid optimization technique called the Chronological-Earth Worm optimization technique (C-EWA) that uses the Gabriel graph model to maintain network connectivity while adjusting power and energy parameters to conduct efficient clustering. The objective function is utilized in the development of this technique. Here, a number of variables including power, connection, mobility, link lifetime, and distance are taken into account by the objective function. 
Study \cite{moaveninejad2005low} provides a distributed and a few centralized algorithms to build a network topology for wireless ad hoc networks in a way that minimizes the maximum interference from links or nodes inside the topology, meeting an expectation like a power, hop, or length. 

Another method for controlling the topology of mobile ad hoc networks is shown in \cite{genda2020topology}. It takes into account the cumulative energy consumption across the update interval and the dynamic placements of surrounding nodes to determine the ideal topology for each topology update interval. 
The suggested method in \cite{pradhan2010adaptive}, called adaptive distributed power management (DISPOW), creates a network topology with an effective connection and a suitable transmit power level in a fully distributed fashion, taking into account the propagation conditions and node density around it. The table \ref{tab:related_work2} provides a synopsis of the key ideas from these papers. 

\begin{table}[h]
	\caption{An overview of some power management studies.}\label{tab:related_work2}%
	\footnotesize
	\begin{tabular}{ |p{0.2cm}||p{1.3cm}|p{1.7cm}|p{1cm}|p{1.6cm}|  }
		\toprule
		paper & model & objectives & application & evaluation metrics \\
		\midrule
		\cite{burkhart2004does} & Local Low Interference Spanner Establisher (LLISE) & to minimize network interference, preserve network connectivity, be network spanners & mobile wireless ad-hoc networks & interference \\
		\hline
		\cite{chaudhry2018optimized} & Optimized Power Control (OPC) & to determine the optimal transmission power of	nodes & MANET & transmit power, network delay, and total energy consumption\\ 
		\hline
		\cite{devika2020power} & Chronological-Earth Worm optimization Algorithm (C-EWA) & to adjust power and energy parameters, maintain network connectivity & MANET & battery power, mobility, throughput, delay and connectivity \\
		\hline
		\cite{moaveninejad2005low} & not named & to minimize the maximum (or average) link (or node) interference & wireless ad hoc networks & node interference and link interference \\
		\hline
		\cite{genda2020topology} & not named & to reduce energy consumption while preserving network connectivity  & MANET & energy consumption and computation time \\
		\hline
		\cite{pradhan2010adaptive} & distributed power management (DISPOW) & to preserve network connectivity and reduce interference & MANET & network connectivity and node power \\
		\botrule
	\end{tabular}
\end{table}

\section{Hamiltonian Approach for Topology Control of the IoT-Based Networks}
\label{sec:methodology}
A network is defined as a collection of items, each of which is referred to as a node and the connection between them as a link. Each network is described according to its structure (nodes and links) and behavior. The word ?behavior? refers to what occurs in the network as a result of interactions between nodes and links. In order to study a network, its structural and dynamic aspects are mostly represented using graph theory and a set of rules that govern node and link behavior \cite{borner2007network}. 

\begin{figure}[!h]
	\centering
	\includegraphics[width=0.5 \textwidth]{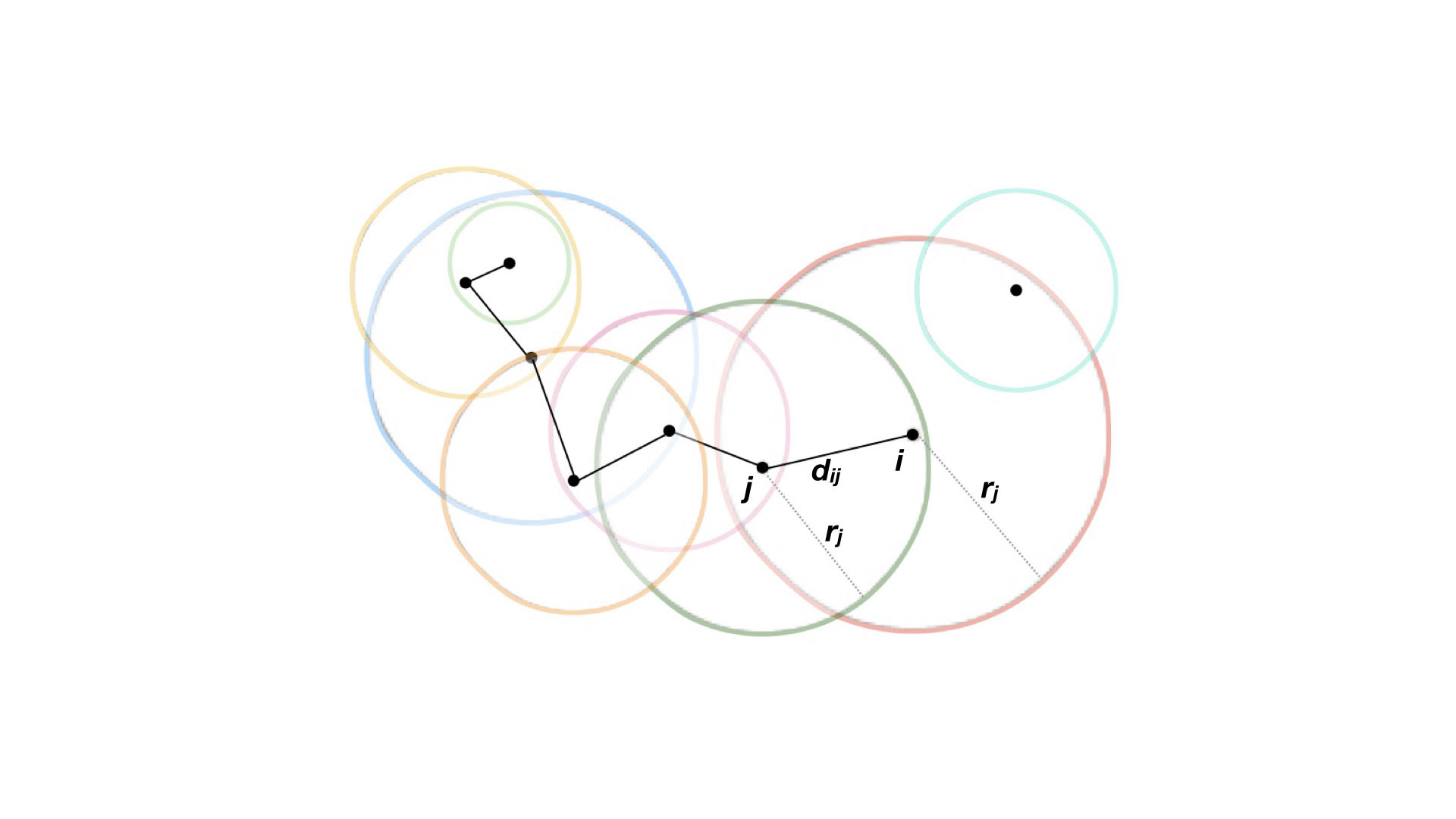}
	\caption{Euclidean distance is the connection metric employed here. In this network, a direct connection or link is created when two arbitrary nodes are within the transmission range of each other.}
	\label{fig:concept_image}
\end{figure}

Nodes in the network {\color{black}(objects in an IoT)} under study have specific spatial positions. In fact we take the geography into account{\color{black}, so our network is categorized as a spatial complex network.} {\color{black} Here, nodes can be made up of different device kinds. Although each of these distinct kinds of nodes may have its own special qualities, they are defined in way that ensures they adhere to the same communication standards and share information in the communication network.} 
In this study, the space is two-dimensional Euclidean space, {\color{black}however the method can be directly applied to the higher dimensions.} The connection metric here is the Euclidean distance. To establish a direct connection or link in this network, two arbitrary nodes must be located inside one another's transmission range (See Figure \ref{fig:concept_image}). 

{\color{black} Nodes' transmission ranges are not considered as being equivalent. It is due to the prevalence of equipment with various transmission ranges in the real world. Additionally, the fact that the nodes' range values differ ensures that there is no correlation between the results and any particular value for the range. Having said so, we take into account a predetermined base value for all the nodes, and we then add a random number generated with normal distribution to this base value to determine the value of each node's range. 
	With this approach, the nodes' ranges will be divers in size but of the same {\color{black}order of} magnitude.} For the channel model, only bidirectional communication links are taken into consideration. {\color{black} This is the only property that we considered for the channel.} Consequently, the mapping {\color{black}network} should be undirected, and all link relations on node pairs and the ensuing adjacency matrix (A) are symmetric. Nodes are labeled, so the adjacency matrix and the {\color{black}network} have a one-to-one relationship.

{\color{black}In order to generate the network, we define and employ a Hamiltonian equation as a cost function that fulfills the desired network characteristics.} The topological features of the network and the mode of occurring communication should be included in the definition of the Hamiltonian function. Two terms of this function, 

\begin{equation}{\label{Ha}}
	\sum_{i}^{} \alpha_i {k_i}^2 + \sum_{i}^{} \beta_i {k_i}^3 ,
\end{equation}

{\color{black} are concerned with the topological distribution of the network and proposed by \emph{Berg} and \emph{Lassig} \cite{berg2002correlated}.} In this formula, $\alpha_i$ and $\beta_i$ are arbitrary coefficients for the node $i$ , and $k_i$ denotes the degree of the node $i$, where $i = 1, 2, ..., N $. {\color{black} $N$ denotes the total number of nodes.} The degree of a node is determined by the number of links that connect it to other nodes in the network \cite{bullmore2009complex, tian2008tale, kim2013coevolution}. 
First term in Eq. \ref{Ha} provides the number of paths on the {\color{black} network} that are two lengths long. It rewards the formation of hubs, i.e., highly connected nodes, which in turn lead to short distances. Actually, this term results in the most compact, star-like configuration possible. The collapse to a star, where the connectivity of the central vertex scales with the size of the {\color{black} network}, is prevented by second term in this equation, which is the regularization term \cite{berg2002correlated}. If the coefficient of this term is positive, the network will develop three-node connections or increase the number of triangles overall. As a result, the network becomes more distributed. 

Regarding the communication qualities of the network, two additional terms,

\begin{equation}{\label{Hb}}
	\sum_{i}^{} \gamma_i {r_i}^2 +  \sum_{i,j(i \neq j)}^{} \lambda_i l_{ij}/d_ij ,
\end{equation}

are taken into consideration for the definition of the Hamiltonian function. {\color{black} Hear} the transmission range of $ith$ wireless device in network is called $r_i$. $d_{ij}$ shows for the Euclidean distance between nodes $i$ and $j$, and $l_{ij}$ represents the state of the connection, {\color{black} i.e. the value of $l_{ij}$ is equal to 1 for any two arbitrary interconnected nodes, while it is equal to 0 otherwise.} First term of Eq. \ref{Hb} {\color{black} concerns the nodes' transmission power.} It is determined based on the \emph{Friis transmission equation}. According to this equation, the transmission power of wireless devices is inversely related to the square of their distance as $p_t \propto 1/r^2 $ \cite{franek2017phasor, alu2010wireless, jamnejad2010study}. 
Considering that devices in ad hoc networks function as both antennas and receivers, they must be able to compensate such power (Energy) loss to maintain communicating. 
On one hand we seek to optimize the energy usage of the network, so each device should have the lowest possible {\color{black}transmission power, although it may result in formation of some disconnected localized subnetworks}. On the other hand network localization is prevented by including the last term of the Eq. \ref{Hb} in the Hamiltonian function.  
When optimizing the system, this term in the Hamiltonian equation forces a few long-distance {\color{black}links to be created}. 

The definition of the Hamiltonian of the wireless network is obtained by putting together the four previously mentioned terms{\color{black}, as follows,}

\begin{equation}{\label{totalH}}
	\begin{aligned}
		H = \sum_{i}^{} \alpha_i {k_i}^2 + \sum_{i}^{} \beta_i {k_i}^3 + \sum_{i}^{} \gamma_i {r_i}^2 + \sum_{i,j(j \neq i)}^{} \lambda_i l_{ij}/d_ij , \\
		= \sum_{i}^{} H_i = \sum_{i}^{} \{ \alpha_i {k_i}^2 + \beta_i {k_i}^3 + \gamma_i {r_i}^2 + \lambda_i \sum_{j(j \neq i)}^{} l_{ij}/d_ij\} ,
	\end{aligned}
\end{equation}

where $H_i$ is the contribution of the node $i$ in the Hamiltonian of the network.
The summation operator, which is included in every expression of the Equation \ref{totalH}, implies the adding up over Hamiltonian function of all the individual nodes. This is according to the definition of the Hamiltonian function that specifies the sum of all the constituent costs of a physical system at any moment. 
{\color{black} The value of the variables $ \alpha, \beta, \gamma,$ and $\lambda $ determine the final steady state of the system. In order to reach this state, we minimize the Hamiltonian function by adjusting the transmission range of the nodes, which causes a few links to be added or removed. 
	The dynamics of the network are set up in such a way that, by selecting the proper coefficients for the Hamiltonian function, the cost and energy (power) usage of the network decrease and its connectivity increases until, at some point, these two parameters attain their equilibrium values.}

{\color{black}Realizing that these four coefficients are scaled relative to one another, the value of $ \alpha$ is regarded as fixed in order to simplify the problem. This value is arbitrary, and all remaining coefficients will be scaled in relation to it. The correct numerical range for $\beta, \gamma,$ and $\lambda$ can be identified by running numerous simulations with a variety of potential values over a broad numerical range, evaluating the results, and then repeating the simulations for additional possible values. In order to reduce the computational costs caused by the high repetition of the simulation, we determined the proper value for two of remaining coefficients, namely $\beta$ and $\lambda$, through trial and error of values of different orders of magnitude.  
	Finally, the value of the sole remaining coefficient, namely $\gamma$ is considered as the variable parameter of the problem. }

In order to consider the problem's general state, we define the parameter T, as a {\color{black} metric to represent the freedom} of each node to affect the status of its communications. We'll see in the remaining parts of this article how some nodes with particular T values {\color{black} prefer to refuse creating a link even if connection metric is satisfied.} 
{\color{black} $T$ and nodes distribution density $(\rho)$ are the two control parameters used in these simulations. The value of $\rho$ conveys an overview of the configuration of nodes in the network domain. Since the network is spatial, its connectivity depends on how nodes are distributed through it. While a sparse network is hardly ever connected, a dense network almost always is.
	Recognizing the $\rho$ and how it relates to N, makes it possible to compute the surface area $(S)$ of network domain as $ \rho = N / S $ for each simulation. 
	When the $\rho$ in the network changes, the sides linear length $(L)$ of the network domain, which is conceived of as a square, will change in proportion to the inverse square root.  }

A Monte-Carlo (MC) link dynamics analysis was employed to conduct the study. 
By employing this technique, researchers can simply repeat or adapt experiments, investigate their hidden quantities, and investigate complex systems \cite{harrison2010introduction}. {\color{black} In this study, Monte Carlo intelligently explores the probability space and obtains the point, where the cost function is minimized.}
{\color{black} 
	As mentioned before, the steady state of the system is determined by Hamiltonian function and the different values of its coefficients determine the different steady states of the system, which can be connected or disconnected. }
The purpose of this study is to influence the network's dynamics in such a way that the system reaches its steady state and adheres to its global minimum {\color{black} of Hamiltonian function} by self-organizing. The system's steady state is defined as the point at which the network reaches a certain, almost constant level of connectivity and oscillates around it with a specific amount of error. In a stable state, the connection status of the links may change, but these local changes have no effect on the network's global connectivity level.

{\color{black} We define and carry out the simulation in three separate phases. 
	In first phase, referred as centralized static network, nodes' control parameters ($T$ and $\rho$) are assumed to have the same value. We minimize the cost function of the network by employing Monte Carlo method to obtain proper values of control parameters to maintain the network in connected state. In this phase, we run the simulation repeatedly with various values for the three parameters $T$, $\rho$, and $\gamma$. From the outcomes, we pick and record the proper $\gamma $ values that get the network to the desired stable state. The value of this coefficient will be employed in subsequent phases in accordance with the relevant control parameters as $\gamma(T, \rho)$ in order to steer the dynamics of the problem in the right direction.
	
	In the next phases, which are static and mobile ad hoc networks, distributed structures are established. Some distinctions are applied to the nodes by allocation of different values for their $T$ and $\rho$. In these ad hoc networks, each node scans its surroundings and estimates the $\rho$ of the network by counting its nearby nodes. With this scenario we assist to increase the accuracy of the problem solution.  
	In these phases, the efficiency of the proposed method is investigated for when there is not a central server and each node tries to minimize its own Hamiltonian function besides selecting the most proper coefficient $\gamma(T, \rho)$ for it. 
	
	In our networks, nodes can alter their connection status with surrounding nodes by varying their transmission power. 
	When nodes in the network become mobile, they move through the transmission range of other nodes to form new links or eliminate any that are no longer essential. The possibility of creating links whose existence is not only unnecessary but also increases energy consumption makes this link removal issue crucial.}
Following, each of these three phases is discussed in more detail.

\subsection{Centralized Static {\color{black} IoT} Network}

The nodes in the network are stationary during this phase. Required information for the nodes to establish and maintain the essential communication in the network is determined by an {\color{black}admin}, who functions as a control center. The Control parameters, i.e. $T, \rho$ and Hamiltonian function's coefficients, i.e. $\alpha, \beta, \gamma,$ and $\lambda $ are part of the required information. Nodes are identically modeled and are in comparable circumstances during this simulation phase. Therefore, the {\color{black}admin}-specified values for the aforementioned parameters of the node are exactly the same. 

We intend to make the network topology as close as feasible to the connected topology by trial and error in assigning various values for the problem parameters. After completing the simulation of this phase, it is possible to identify the proper $\gamma(T, \rho)$ values that bring the network to connectivity $\geq$ 95\% and use them in the next phases. Therefore, this phase could also be referred to as the \emph{calibration phase}. 

Initially a number of nodes are distributed randomly in a specified space as the wireless IoT devices. The nodes that are within transmission range of each another are linked after the nodes are located. The initial network topology has developed at this point. Because nodes are stationary, the network connectivity can only be changed by adjusting the nodes' transmission range. {\color{black}$\Delta r$ must meet the requirements of Eq. \ref{deltaR}. In this equation, $s$ and $s'$ represent the different arbitrary Monte Carlo steps.} For changing the range of nodes gradually, the Monte Carlo method is employed. 

Algorithm  \ref{alg:one} is the recommended route of action for presented strategy. According to this Algorithm, for $T=0$, only the energy minimization condition works. However, for $T > 0 $, the system is free to experience {\color{black} temporary} positive cost changes as well. After the system has stabilized and the relation between its $T$, $\rho$ and $\gamma$ parameter values has been documented, new values are given to these parameters. A new network topology is then constructed, and the previously indicated procedures are repeated. For multiple values of $T$ and $\rho$ in the network, all the proper $\gamma(T, \rho)$ have been discovered up to the end of the simulation.

{\color{black}
	\begin{equation} \label{deltaR}
		\begin{cases}
			<\Delta r_i> = 0 \\
			< \Delta r_i(s) \Delta r_j(s') > = 2 \delta_{ij} \delta (s-s') \\
		\end{cases}    
	\end{equation}
}

\begin{algorithm}
	\caption{Monte Carlo algorithm for topology control of the centralized static IoT network.}\label{alg:one}
	\begin{algorithmic}[1]
		\small
		\Require initial topology
		\Ensure final topology
		\For{$k \gets 1$ to \textit{Number of  Monte Carlo Steps}}
		\State A single node $i$ is chosen at random.
		\State The $r_i$ is modified by adding a normaly distributed random number called $\Delta r_i$, in  the form ${r_i}^{new} = r_i + \Delta r_i $.
		\State Connections in network are reconfigured and $A$ is updated considering the ${r_i}^{new}$.
		\State $ \Delta H = H^k - H^{k-1} $ is calculated using Eq. \ref{totalH}.
		\If{\textit{$ \Delta H < 0$}}
		\State Modifications are approved.
		\Else
		\If{\textit{ $ e^{{\color{black} - \Delta H/T} } > $ a uniformly distributed random number in [0, 1]}  }
		\State Modifications are approved.
		\Else
		\State $r_i = {r_i}^{k-1}$ 
		\State $A = {A}^{k-1}$
		\EndIf
		\EndIf
		\EndFor
	\end{algorithmic}
\end{algorithm}

\subsection{Ad Hoc {\color{black} IoT} Network}
In the second phase, we take things a step further and think about modeling a distributed network. Nodes in this network are still static.
{\color{black} To make the network more realistic, varied $T$ values are considered for the nodes. In our ad hoc networks, the admin goes away and nodes try to estimate and find the most proper values for their remaining parameters. 
	The constant transmission of information between all the nodes is not intended in order to avoid data flooding in the network. So, the status of the nodes in the network is unknown to each other. Instead, each node can estimate the $\rho$ of the network by locally scanning its surroundings and realizing the nearby nodes distribution density within a specific distance.
	Therefore, their assumptions of the $\rho$ may differ. It can help to increase the accuracy in the $\gamma (T, \rho)$ selection. 
	For possible values of $T_i$ and $\rho_i$, the most proper $ \gamma_i $ values for leading the network to the aimed topology, are acquired in the preceding phase. 
}As a result, the findings so far help the network to self-organize more efficiently by better building each node's Hamiltonian function. 

The distributed nature of the network {\color{black} in absence of any admins} causes some differences in the recommended Monte Carlo algorithm for this phase. We refer to it as {\color{black} \textit{distributed Monte Carlo} algorithm}, Algorithm \ref{alg:two}. Each distributed Monte Carlo step also includes transmission range adjustments, but not only for one randomly chosen node; instead, transmission range modifications are possible for all nodes with a given probability. We set this probability to be equal to $1/N$, so that the results of this phase can be statistically comparable to the results of the previous phase. Each node independently makes these local adjustments to its $r$. 

In addition, at this stage, to provide more freedom of action to the nodes by their $T$, the ability to influence the creation or non-creation of new links is considered for them in some situations that are affected by transmission range modifications of other nodes. 
So, a minor adjustment must be made to the connection metric for the distributed networks. In this case, if two nodes depart each other's transmission range for whatever reason, the link between them will be broken, just like previous phase. However, if two nodes become included in each one's transmission ranges, a link may not necessarily be formed between them, thus it needs to look into the \textit{local cost condition} for those nodes (See Algorithm \ref{alg:dcm}).
The network self-organizes as a consequence of the nodes' independence.

\begin{algorithm}
	\caption{Distributed connection metric algorithm for node $i$.}\label{alg:dcm}
	\begin{algorithmic}[1]
		\small
		\Require nodes $i$ and $j$
		\Ensure $A_{ij}$
		\State $d_{ij}$ is calculated.
		\If{$ d_{ij} \leq minimum(r_i, r_j) $}
		\If{ \textit{ $ e^{ -\Delta H_j/T } > $ a uniformly distributed random number in [0, 1]}}
		\State $A_{ij} = 1$
		\Else 
		\State $A_{ij} = 0 $
		\EndIf
		\Else
		\State $A_{ij} = 0 $
		\EndIf
	\end{algorithmic}
\end{algorithm}

\begin{algorithm}
	\caption{Monte Carlo algorithm for topology control of the Ad Hoc IoT Network.}\label{alg:two}
	\begin{algorithmic}[1]
		\small
		\Require initial topology
		\Ensure final topology
		\For{\textit{ $k \gets 1$ to  Number of distributed Monte Carlo Steps}}
		\For{\textit{ $i \gets 1$ to  N}}
		\If{\textit{ a uniformly distributed random number in [0, 1] $ < 1/N $}}
		\State The $r_i$ is modified by adding a normaly distributed random number called $\Delta r_i$, in the form ${r_i}^{new} = r_i + \Delta r_i $.
		\State Connections in network are reconfigured and $A$ is updated considering the ${r_i}^{new}$ and distributed connection metric (Algorithm \ref{alg:dcm}).
		\State $ \Delta H = H^k - H^{k-1} $ is calculated using Eq. \ref{totalH}
		\If{\textit{$ \Delta H < 0$}}
		\State Modifications are approved.
		\Else 
		\If{\textit{$ e^{ - \Delta H/T} > $ a uniformly distributed random number in [0, 1]}}
		\State Modifications are approved.
		\Else
		\State $r_i = {r_i}^{k-1}$ 
		\State $A = {A}^{k-1}$
		\EndIf
		\EndIf
		\EndIf
		\EndFor
		\EndFor
	\end{algorithmic}
\end{algorithm}

\subsection{Mobile Ad Hoc {\color{black} IoT} Network}

Simulating various IoT networks is possible using the ad hoc model described in the previous phase. However, MANETs are the vast amount of application networks in the actual world. It implies that devices in the network, or at least some of them, have the ability to move. At this level, we consider the network nodes to be mobile and run the simulation in a manner that is similar to previous phases. we are going to demonstrate that the proposed algorithm for distributed ad hoc networks are applicable even for networks constituting with mobile nodes. 
The notion of time should also be taken into account when considering movement of nodes. In this simulation, only one Monte Carlo step is carried out at each time step for sake of simplicity.	

Providing the nodes with the ability to move is defined as assigning them an initial {\color{black}velocity} and a motion model. Initial velocity is generated at random using a uniform distribution, across a predetermined interval. For mobility of the nodes, there are various models that can be taken into consideration. {\color{black} The \emph{random walk model} is employed in our simulation.  
	In this model, an object's next position is wholly independent of its current position \cite{lawler2010random}.} 
It is employed to increase the unpredictability in the nodes' mobility behavior to demonstrate that the final topology is independent of the nodes' motion model. The nodes' motion equation is as follows:

{\color{black}
	
	\begin{equation} \label{EqMo}
		\begin{cases}
			v(t) = \eta(t) \\
			x(t) = v(t) + x(t-1) \\
			y(t) = v(t) + y(t-1) \\
		\end{cases}
	\end{equation}
}

{\color{black}In Eq. \ref{EqMo}, $\eta (t)$ is a normally distributed random number within the predetermined velocity interval.} Boundary conditions are considered for the simulation while being aware of the network environment's area restriction. Due to the nodes' continuous roaming, new links are constantly created while some existing links are broken. Accordingly, the network will be more dynamic than it was in the previous phases. 
The estimated value of $\rho$ by each node will fluctuate as an outcome of the neighborhood modifications. Each node should scan its surrounding environment and update its $\rho_i$ in each {\color{black} time} step. The coefficients of the Hamiltonian function for each node are revised each time the value of $\rho_i$ is scanned and updated; This includes using first phase outcomes. The network can reach its steady state when the Hamiltonian function is adjusted continually by tracking the nodes' movement. After the network reaches stability, this dynamic of local adding and dismissing links will continue, but with much less impact on the global network characteristics, such as connectivity.

\section{Analysis of Results and Discussion}
\label{sec:results}
\subsection{Simulation Evaluation Metrics}

Metrics that are used to analyze and evaluate the performance of the proposed model, are briefly covered in this subsection.

\begin{enumerate}
	\item \textbf{Cost} is one of the primary quantities examined in this study, whose value can be obtained by evaluating the Hamiltonian (H) function of the network. By minimizing its value, a stable network topology is achieved. The network reaches its stable state when the cost falls off as the network dynamics advance, saturates at a specific value, and then oscillate around this value within a certain constrained range. This means that moving forward, there will be zero or negligible cost fluctuations between successive monte carlo steps ($\Delta H \simeq 0 $). A prerequisite for the success of the proposed model is that the coefficients of the Hamiltonian function for each network and its related dynamics should be chosen so that such changes comes at cost.
	\item \textbf{Connectivity} of the network can be measured by employing the concept of a Network's giant component. In many networks, the existing components gradually join to one another as the network evolves. In some instances, a considerable fraction of the nodes are eventually interconnected to form a single, massive component known as a giant component \cite{dereich2013random, ta2009giant, dorogovtsev2001giant, tishby2018revealing} . If the network dynamics in a particular simulation point away from the formation of the giant component, it signifies that the simulation's chosen parameters lacked the necessary synchronization for the solution of our problem. The percentage of network nodes that belong to giant component once the network reaches stability is recognized as a good indicator of the network connectivity. In our simulation, connectivity is desirable when the size of the emerged giant component reaches at least $95\%$ the size of the network and remains constant at or around this value.
	\item \textbf{Energy} of the network is $r^2$-proportional according to Eq. \ref{Hb}. In our analysis, the energy is measured and compared based on this premise and there is no need to compute and know the net quantity of network energy. Through the computation and comparison of $r^2$, one can assess the energy consumption of the network under various conditions, as well as identify the optimal transmission power $r$ for the devices while preserving network connectivity. The proposed method claims that once the network stabilizes, the average transmission radius of the devices should achieve an acceptable and reasonable fraction of the network dimensions, and the energy consumption should be lower than under the basic conditions.
	\item \textbf{Communication stability ($\tau$)} is not a factor that may be used to assess if the objective of the study has been met. Instead, it is a metric for evaluating how stable communications or {\color{black}paths} are within a network. This metric is employed when alternative tuples of parameters $(T, \rho, \gamma)$ assist the network reach its objective; in such a case, the tuple that preferably has better communication stability is chosen. Within the specific time frame of the entire simulation time, a link could exist between any two arbitrary nodes in the network. This link could exist in a fixed and unchanging form for the duration of this particular time period, or it could connect and disconnect numerous times as the network develops. A link's status is regarded as being more stable if it is more consistently present in the network and experiences fewer connections and interruptions. The communication stability ($\tau$) is presented as the average stability of all potential links in the network as Eq. \ref{eq:tau}. In this equation $M$ stands for the number of all potential links in the network and it is equal to $ M = N*(N-1)/2 $. $s$ shows the number of Monte Carlo steps executed up to that moment. $P_{ij}$ is the number of Monte Carlo steps, which had $A_{ij} = 1$ to that moment and $Q_{ij}$ is the number of steps in which the value of $A_{ij}$ is changed.

	\begin{equation}{\label{eq:tau}}
		\tau = 1/(sM) \sum_{ij} (P_{ij}/Q_{ij})
	\end{equation}

\end{enumerate}

\subsection{Results}

This subsection describes the outcomes of the Hamiltonian approach for topology control of the IoT networks. We start by discussing the values that were determined for the control parameters of our simulation case. Number of nodes (N) in all three phases  of the simulation is regarded as being identical, which is set at 100. We consider these 100 nodes as belonging to one of three different device categories. 
For transmission range of nodes (r), a predetermined base value is considered that is about $0.05 L$ in the centralized static network simulation. 
In each monte carlo step, where it is necessary to modify the node's transmission range, the value for this range change $(\Delta r)$ is chosen by a normally distributed random number generator.

In the simulation of the first phase, the value of T starts at 0, indicating that there is no disorder in the network, and rises to a certain value, which in this case is equal to 1000, in order to study the impact of the $T$ with values in this numerical range on the behavior of the nodes and the dynamics of the network.
In distributed phases, different value for $T$ of each node is considered, generated at random with uniform distribution within the given interval. 
The value of network's node distribution density $(\rho)$ in our simulation varies from almost $0$ to roughly $0.1$, which is a substantial amount for the node density in an ad hoc network in the actual world. In distributed networks, each node is responsible for scanning its surroundings and estimating the density of nearby nodes. Coefficients of the Hamiltonian function $( \alpha, \beta, \gamma, \lambda )$ are other parameters of the problem. $\alpha$  is always set to $-0.5$. A negative value is chosen because it aids in achieving the global minimum by lowering the Hamiltonian function's output. The values of $\beta = 0.3$ and $\lambda = -1000$ have been found through our experimentation to be appropriate. Several simulations indicated that the network become steady only when the numerical value of $\lambda$ is grater than a threshold limit and it is also negative.
We take into account the value of $\gamma$ starting at 0 and raising it to 100 for the given values of the $T$ and $\rho$. Number of Monte Carlo steps is a parameter whose optimal value is established via experimentation and tracking the outcomes. 

The value of this parameter is adjusted for each simulation in a way that makes it evident from the results whether the network become stable or not. Here, number of Monte Carlo steps is set to 15000 and in some cases 30000. 
In order to analyze and evaluate the impact of proposed topology control algorithm, we will compare its outcomes and dynamics with those of the network with base dynamic. Our base dynamic is precisely the same as that the proposed method analyzed; the only differences are that the dynamics established by varying the radius and placement of the devices are implemented without any additional constraints, and the control topology algorithm is not applied to the network. 
C++ programming language is used to simulate the desired network environment and its dynamics, and the python environment is used to plot the diagrams. 

There are a few main reasons why we decided to use C++ to develop our simulation environment. Given the computationally demanding tasks required in real-world network simulations, its high performance and efficient memory management are essential. Complex network model simulation can be implemented more easily thanks to its object-oriented capabilities, which allow for modular and scalable code development. The extensive standard libraries and fine-grained control over hardware resources that C++ offers further improve the accuracy and adaptability needed for our particular simulation requirements. In comparison to alternative network simulators and programming languages, C++ enables us to carefully design and optimize our simulation environment to our research needs, promising reliable and accurate results.

\subsubsection{Centralized Static IoT Network}

\begin{figure}[!h]
	\centering
	\includegraphics[width=0.4 \textwidth]{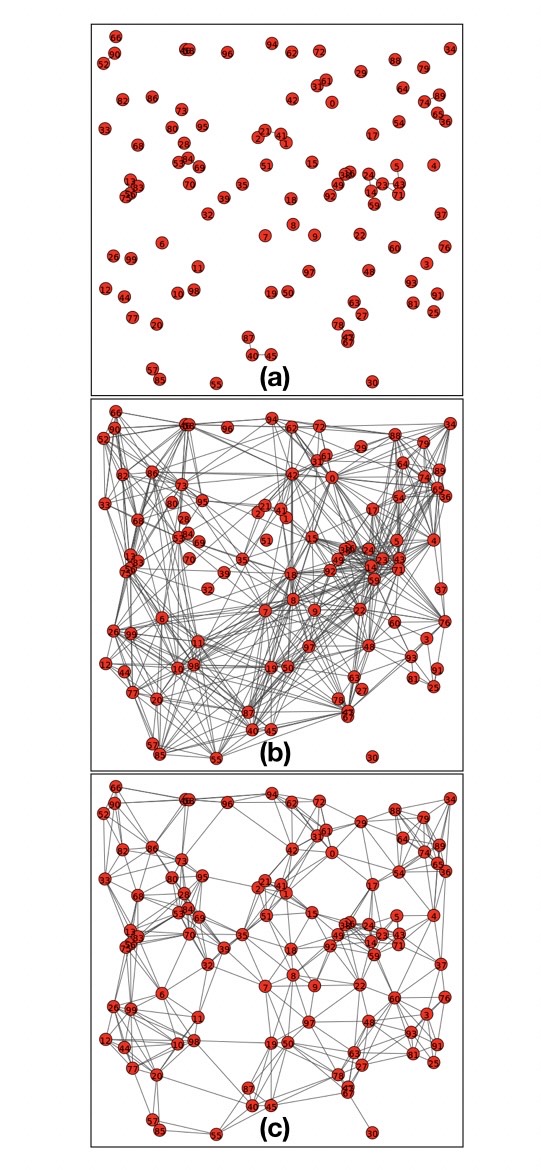}
	\caption{(a) represents the initial spatial network constructed before any modifications at step 0. (b) represents the network topology following the execution of 15000 Monte Carlo steps with basic algorithm. (c) is the same network after execution of 15000 Monte Carlo steps with topology control algorithm. These snapshots are from a centralized static network simulation with the control parameters $\rho = 0.05, T = 100$, and $\gamma = 1$.}
	\label{fig:global_snapshots}
\end{figure}

In order to find most proper parameter values for the distributed networks, we will examine and discuss the results of the centralized static network simulation in this part. 
Three distinct visualizations of our simulated centralized network in different states and times are displayed in figure \ref{fig:global_snapshots}. Figure \ref{fig:global_snapshots}. (a) depicts the network's post-formation state prior to the beginning of its dynamics. It is evident from this picture that the network is mostly isolated, with few connections, as a result of the relatively low average transmission radius. We use the basic algorithm and the topology control algorithm to execute 15,000 Monte Carlo steps of simulation. Figures \ref{fig:global_snapshots}. (b) and \ref{fig:global_snapshots}. (c) display the network snapshots that were taken at the conclusion of these simulations. They can be compared to demonstrate how the suggested topology control technique dramatically lowers the number of links in the network while preserving its connectivity. Definitely, the network's energy consumption will drop significantly if this many links are taken out. Figure \ref{fig:global_snapshots} has well demonstrated the changes before and after the simulated dynamics. As shown in figure \ref{fig:global_snapshots}. (c), with much fewer links than all possible links, the network has achieved stability and is connected.

\begin{figure}[!h]
	\centering
	\includegraphics[width=0.5 \textwidth]{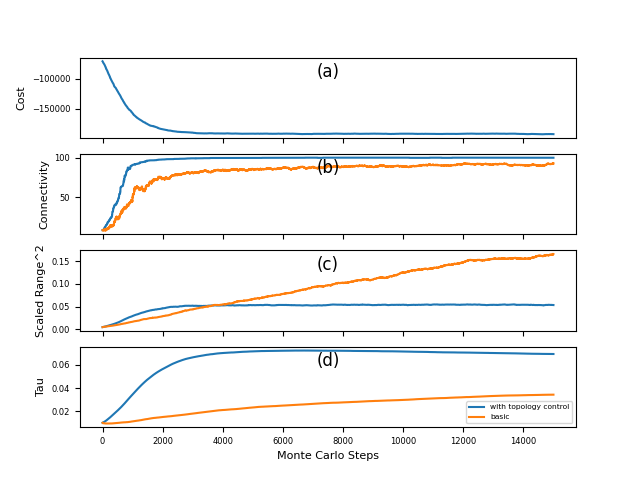}
	\caption{Variation diagrams for (a) cost (b) connectivity (c) average scaled ranges$^2$ (d) average Tau over Monte Carlo steps is demonstrated for a centralized static network with the control parameters $\rho = 0.05, T = 100 $, and $\gamma = 1$. These diagrams represent the average of ten statistical ensembles of the intended network.}
	\label{fig:global_compare_four}
\end{figure}

To evaluate the performance of the Hamiltonian topology control approach, we consider the variations in the aforementioned four evaluation metrics during the simulation. Figure \ref{fig:global_compare_four} displays the evolution of these four quantities for a centralized static network with the control parameters $\rho$ = 0.05, $T$ = 100, and $\gamma$ = 1. The diagrams related with the applied topology control approach and the basic dynamic diagrams have been drawn together in order to compare and gain a better understanding of its impact. This figure's part (a) illustrates how the cost value changed throughout the simulation. Since the concepts of cost and Hamiltonian function are not defined for networks with basic dynamics, the diagram of this quantity is only generated for the topology under control. As a consequence of the correct and well-coordinated selection of the values of the problem's control parameters, the system has attained its steady state, as evidenced by the cost's falling trend in the early steps and its almost absolute value after roughly 3000 Monte Carlo steps.

Observing Figure \ref{fig:global_compare_four} closely, we notice that the graphs pertaining to connectivity, scaled range$^2$, and $\tau$ all clearly show the system in its steady state. These graphs similarly attain and hold nearly constant values at the same point that the cost stabilizes. Part (b) of this figure makes it evident that, in steady state, the network connectivity level has hit 100\% and has remained there the whole simulation. This is the ideal connectivity rate. In this case, the giant component contains all the nodes in steady state of the network, which fulfills our goal of 95 \% connectivity. A diagram for evolution of the average scaled range in terms of the steps, is presented in Figure \ref{fig:global_compare_four}.(c). Scaled range is defined as the ratio of transmission range over linear length of network domain.  This figure's (c) part shows that, in the system's stable state, the average square of the scaled range equals 0.05. In other words, the average transmission range needed to achieve 100\% connectivity in this specific network is equivalent to 0.2 of its dimensions. 

The adoption of the Hamiltonian topology control technique has enabled the network to reach a higher level of connectivity with significantly reduced energy consumption during its dynamics, as we can see by comparing the graphs associated to the network with basic dynamics and with topology under control. Furthermore, there is a noticeable increase in link stability in the controlled network.
These outcomes indicate that for the specified parameter values, the equilibrium state and the global minimum point of the system is achieved. At this point, the network is robust to small local changes and is not easily affected. 

Due to issues such as energy resource limitations or environmental occurrences, devices in the network may fail or be attacked. Such misfortunes can easily impair the network's connectivity and communications, and may even severely disrupt it. We investigate and assess the robustness of the controlled network topology against attacks and failures by modeling this procedure. In order to achieve this, we turn of a specific number of nodes every 6000 steps (starting from the beginning of the simulation) during the 30000 Monte Carlo steps of the simulation. The turned off nodes are randomly selected, so a node may be selected more than once in one step. This turn off is described as zeroing the transmission power and cutting all connections between the node in concern and the network. 

\begin{figure}[!h]
	\centering
	\includegraphics[width=0.5 \textwidth]{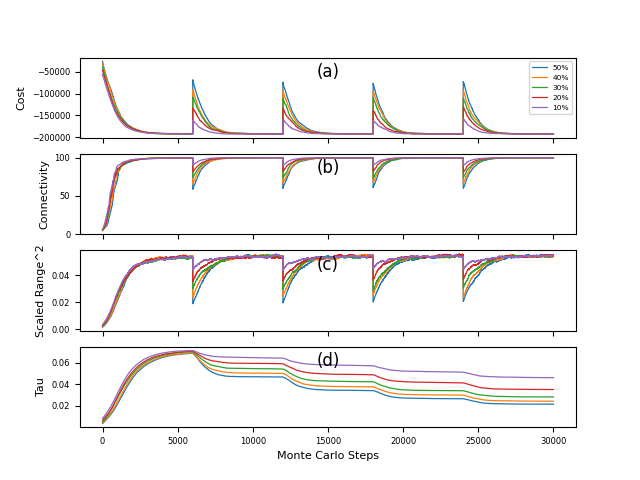}
	\caption{Variation diagrams for (a) cost (b) connectivity (c) average scaled ranges$^2$ (d) average Tau over Monte Carlo steps is demonstrated for a centralized static network facing node failures. The network is created with the control parameters $\rho = 0.05, T = 100 $, and $\gamma = 1$. The diagram's colors represent the different node percentages of failed nodes. These diagrams represent the average of ten statistical ensembles of the intended network.}
	\label{fig:global_compare_four_failure}
\end{figure}

As the simulation progresses, turned off nodes can return to network interactions. In this case, at different points of the simulation, the outcomes will be comparable. Figure \ref{fig:global_compare_four_failure} illustrates variation diagrams of this procedure for turning off up to 10\%, 20\%, 30\%, 40\%, and 50\% of total network nodes. From the diagrams of this figure, it is clear that even despite turning off a significant fraction of nodes, the Hamiltonian algorithm can bring the network back to a stable state and a topology with the desired connectivity.  

\begin{figure}[!h]
	\centering
	\includegraphics[width=0.5 \textwidth]{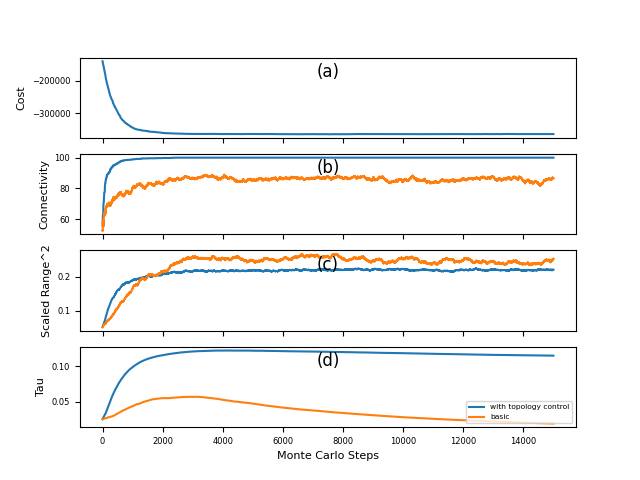}
	\caption{Variation diagrams for (a) cost (b) connectivity (c) average scaled ranges$^2$ (d) average Tau over Monte Carlo steps is demonstrated for a 3D centralized static network with the control parameters $\rho = 0.05, T = 100 $, and $\gamma = 1$. These diagrams represent the average of ten statistical ensembles of the intended network.}
	\label{fig:global_compare_four_3d}
\end{figure}

Our recommended Hamiltonian approach is a general strategy that can be expanded  to handle more particular networks. As far as 2-dimensional centralized static networks are concerned, this model has fulfilled its defined objectives. In the following, we explore the idea of generalizing the current method to a 3D network in order to demonstrate its capabilities. To accomplish this, we create diagrams showing the development of evaluation metrics for simulating the basic and controlled topology dynamics of a 3D network that is exactly like the prior network in terms of all parameters. The interpretation of these diagrams, which are displayed in figure \ref{fig:global_compare_four_3d}, is quite similar to that of the 2D cases diagrams. This graphic illustrates how the topology control method was applied to a 3D centralized static network, enabling the network to achieve the desired maximum connectivity while using less energy and improving the stability of the network connections ($\tau$). As seen in diagram (c) of this figure, an average transmission range of $\sqrt{0.2}$ network dimensions is needed to establish and maintain 100\% connectivity. This number is greater than its corresponding quantity in 2D case, due to the different network dimensions in 2D and 3D configurations for a same number of nodes with the same distribution density. 

\begin{figure}[ht]
	\centering
	\begin{minipage}[b]{\linewidth}
		\includegraphics [width=\linewidth] {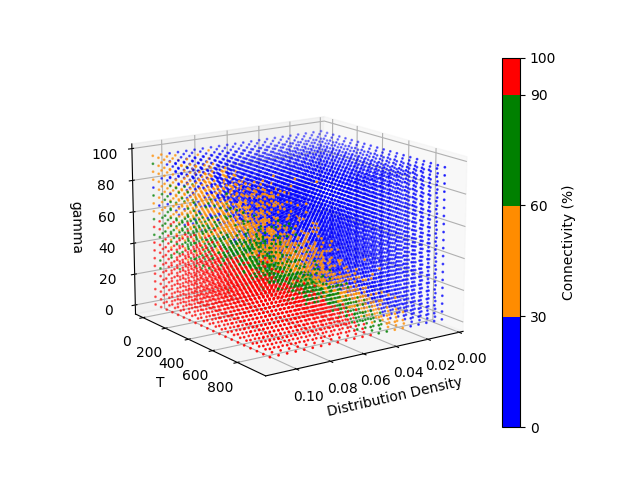}
		\caption{Phase space of the system is illustrated in this diagram. different colors demonstrate different levels of network connectivity. }
		\label{fig:4dconnectivity}
	\end{minipage}
	\quad
	\begin{minipage}[b]{\linewidth}
		\includegraphics [width=\linewidth] {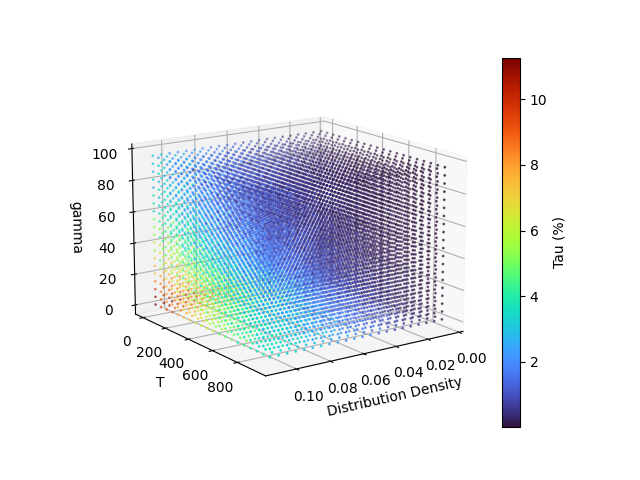}
		\caption{Values of $\tau$ for each simulation with distinct control parameters are demonstrated with different colors in this diagram.}
		\label{fig:4dtau}
	\end{minipage}
	\caption{Phase spaces for Centralized Static IoT Network.}
	\label{fig:phasespaces}
\end{figure}

The parameter space of the simulated system is shown at specific intervals in the Figure \ref{fig:phasespaces}. Each point in this figures depicts individual simulations with distinct parameter settings. 
Figure \ref{fig:4dconnectivity}, which is a 4D diagram, provides us with an overview of the network's connected and disconnected phases based on the values of the three control parameters for this problem, i.e. $T, d,$ and $\gamma$. Different colors indicate the connectivity percentage. The network is connected to just a section of the parameter space, as shown in the figure, and we must pick parameter values from this region for the network to be in its connected phase. According to this figure, it is possible to understand the variation of connectivity in terms of $\rho$ for specific values of two other control parameters. This figure also makes it evident how connectivity has changed in comparison to $T$ and $\gamma$.
Figure \ref{fig:4dtau} is similar to Fig. \ref{fig:4dconnectivity}, with the difference that its color informs us of the $\tau$ parameter. It means that the greater a simulation' $\tau$ is, the more relatively stable its links are. A table is also constructed in this phase by taking into account the control parameter values and output data. With knowledge of the $\rho$ and $T$ of the nodes, the table will be used by each node in distributed phases to determine the most suitable $\gamma$ for their Hamiltonian function.

\subsubsection{Ad Hoc IoT Network}

\begin{figure}[!h]
	\centering
	\includegraphics[width=0.4 \textwidth]{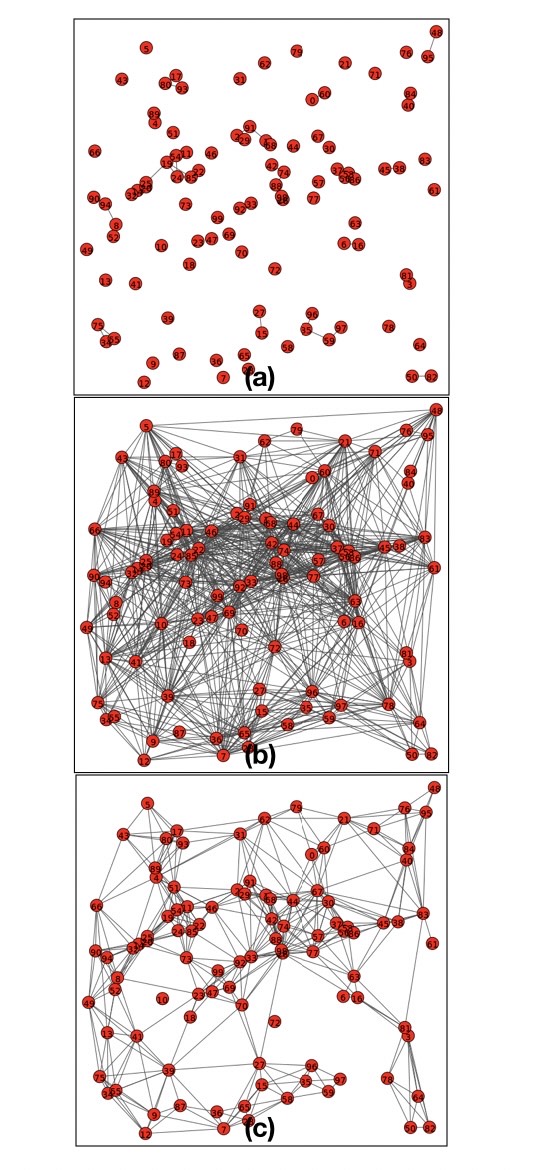}
	\caption{(a) represents the initial spatial network constructed before any modifications topology at step 0. (b) represents the network topology following the execution of 15000 Monte Carlo steps with basic algorithm, and (c) shows final network topology after execution of 15000 Monte Carlo steps with topology control algorithm.  These are snapshots from a simulation of a static ad hoc network with the network distribution density $\rho = 0.05$.}
	\label{fig:local_snapshots}
\end{figure}

\begin{figure}[!h]
	\centering
	\includegraphics[width=0.5 \textwidth]{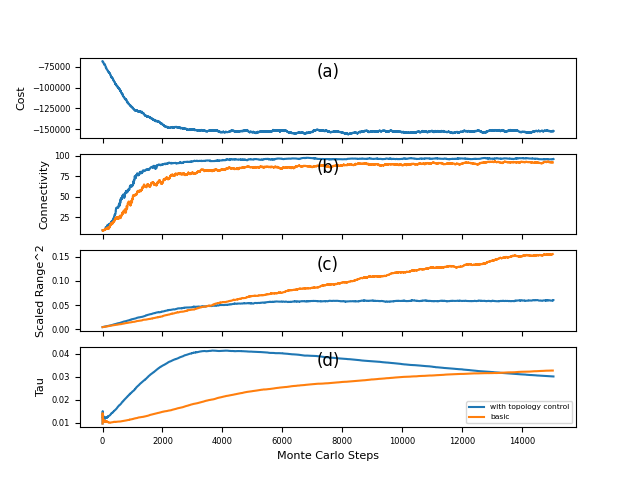}
	\caption{Variation diagrams for (a) cost (b) connectivity (c) average scaled ranges$^2$ (d) average Tau over Monte Carlo steps is demonstrated for an ad hoc network with the control parameter $\rho = 0.05$. These diagrams represent the average of ten statistical ensembles of the intended network.}
	\label{fig:local_compare_four1}
\end{figure}

After obtaining the appropriate Hamiltonian coefficients and dependency table between the control parameters of the system, it is time to simulate an ad hoc network. Figure \ref{fig:local_snapshots} shows snapshots of the simulated ad hoc network topology taken at the beginning (See Fig. \ref{fig:local_snapshots}. (a)) and end of the basic (Fig. \ref{fig:local_snapshots}. (b)) and topology control simulation (Fig. \ref{fig:local_snapshots}. (c)). According to this, network's goal seems to be achieved, although its links appear to be less efficient than in previous phase.
Figure \ref{fig:local_compare_four1} shows the changes of cost, connectivity, average scaled range$^2$, and $\tau$ for the simulated network. This figure leads to the conclusion that the nodes can autonomously adjust their coefficients in the distributed phase using the prepared table and that the first phase's coefficients are coordinated in the distributed phase. 

Similar to the previous phase, the stability happens approximately after roughly 3000 Monte Carlo steps.
The Energy used and average scaled transmission range (See Fig \ref{fig:local_compare_four1}. (c)) is approximately similar during this phase's stability to it was during the preceding phase. 
It is possible to say that the more fluctuations in this phase after obtaining stability is due to the fact that individual nodes are optimized using the local information that they have access to, rather than being updated with information from the entire network.

\begin{figure}[!h]
	\centering
	\includegraphics[width=0.5 \textwidth]{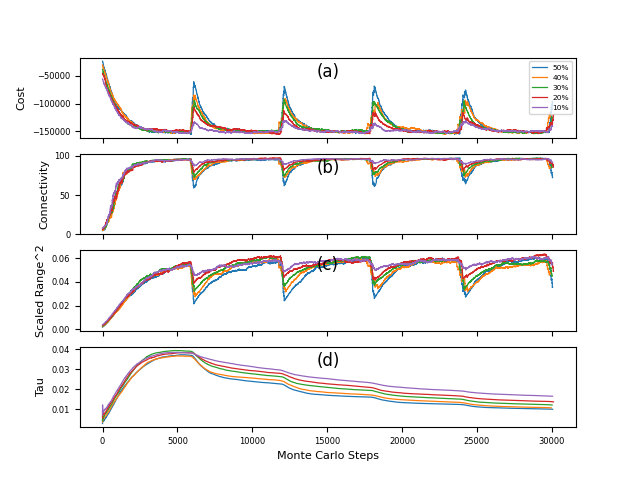}
	\caption{Variation diagrams for (a) cost (b) connectivity (c) average scaled ranges$^2$ (d) average Tau over Monte Carlo steps is demonstrated for an ad hoc network facing node failures. The network is created with the node distribution density $\rho = 0.05$. The diagram's colors represent the different node percentages of failed nodes. These diagrams represent the average of ten statistical ensembles of the intended network.}
	\label{fig:local_compare_four_failure}
\end{figure}

\begin{figure}[!h]
	\centering
	\includegraphics[width=0.5 \textwidth]{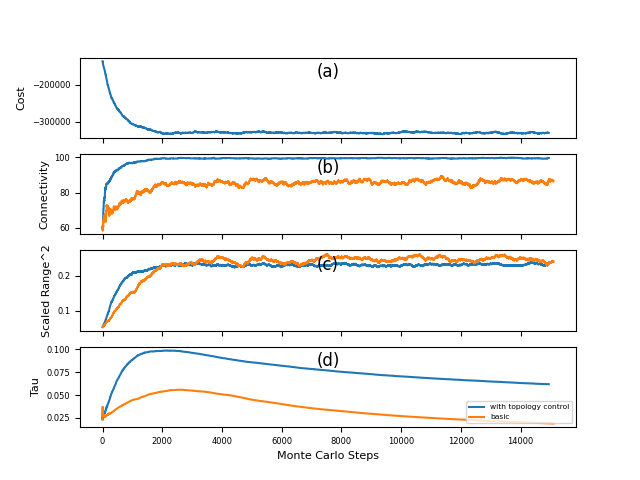}
	\caption{Variation diagrams for (a) cost (b) connectivity (c) average scaled ranges$^2$ (d) average Tau over Monte Carlo steps is demonstrated for a 3D ad hoc network with the node distribution density $\rho = 0.05$. These diagrams represent the average of ten statistical ensembles of the intended network.}
	\label{fig:local_compare_four_3d}
\end{figure}

Figure \ref{fig:local_compare_four_failure} illustrates variation diagrams of node failures for about 10\%, 20\%, 30\%, 40\%, and 50\% of all nodes. The diagrams of this figure proves that even despite turning off a large percentage of nodes, the applied topology control algorithm can bring the ad hoc network back to a stable state which has the desired connectivity. With identical parameter choices, we simulate the 3D ad hoc network and assess it using the problem's metrics. Similar to the previous instances, Figure \ref{fig:local_compare_four_3d}'s diagrams of these metrics shows how well the suggested model accomplished its objectives.

\subsubsection{Mobile Ad Hoc IoT Network}

\begin{figure}[!h]
	\centering
	\includegraphics[width=0.4 \textwidth]{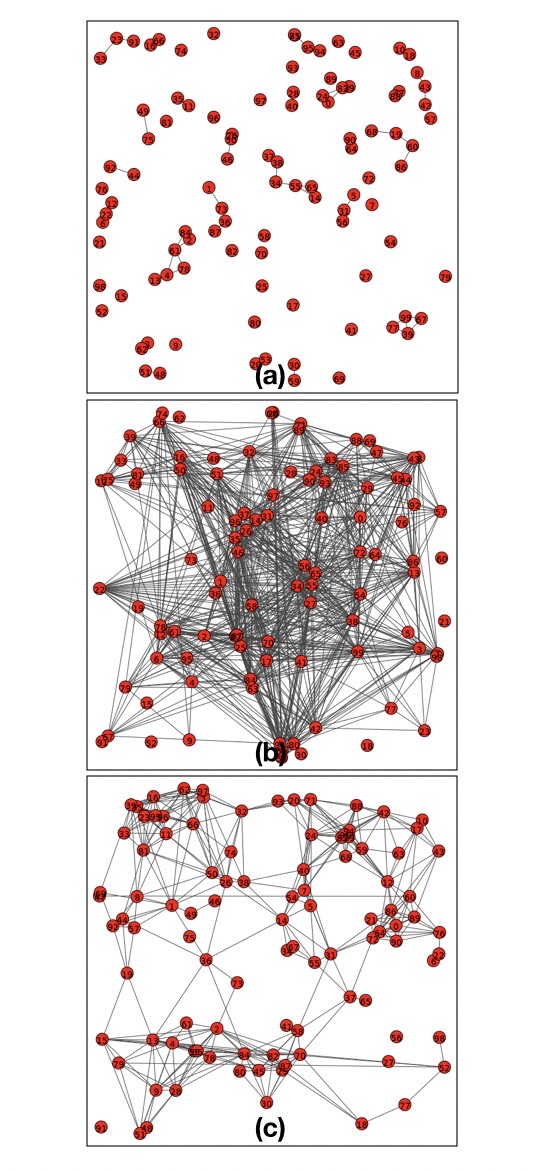}
	\caption{(a) represents the initial spatial network constructed before any modifications topology at step 0. (b) represents the network topology following the execution of 15000 Monte Carlo steps with basic algorithm, and (c) illustrates final network topology after execution of 15000 Monte Carlo steps with topology control algorithm.  These are snapshots from a simulation of a mobile ad hoc network with the network distribution density $\rho = 0.05$.}
	\label{fig:movement_snapshots}
\end{figure}

\begin{figure}[!h] 
	\centering
	\includegraphics[width=0.5 \textwidth]{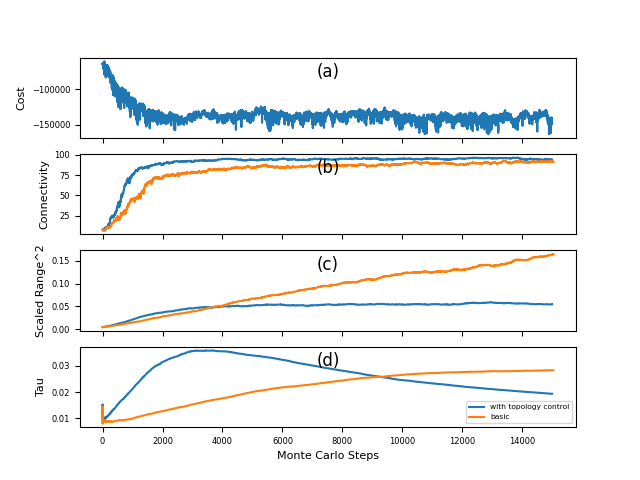}
	\caption{Variation diagrams for (a) cost (b) connectivity (c) average scaled ranges$^2$ (d) average Tau over Monte Carlo steps is demonstrated for a mobile ad hoc network with the control parameter $\rho = 0.05$. These diagrams represent the average of ten statistical ensembles of the intended network.}
	\label{fig:movement_compare_four}
\end{figure}

It's time to simulate the topology control of the mobile ad hoc network after simulating the topology control of the ad hoc network in previous phase and seeing how the results were satisfactory. With the exception of the node's mobility, the network under study is remain the same as the network from the previous phase. The nodes' maximum movement velocity is regarded as 0.3.
The figure \ref{fig:movement_snapshots} displays the topology of the network after its construction as well as at the conclusion of its basic and also topology controlled dynamics. The effect of the applied topology control algorithm on a MANET can be estimated by comparing Fig. \ref{fig:movement_snapshots}. (b) and Fig. \ref{fig:movement_snapshots}. (c).
Variations of cost, connectivity, average scaled range$^2$, and $\tau$ for the simulated MANET are depicted in Figure \ref{fig:movement_compare_four}. The plots in this figure demonstrate how well the developed algorithm and the derived coefficients function in this phase. The network finds stability after roughly 3000 Monte Carlo steps, just like in the previous instances. Results shown in this figure differ a little from those of the preceding phases due to its massive fluctuations, which seem normal given the mobility of its component nodes. The resulting connectivity appears to be satisfactory despite these oscillations.

\begin{figure}[!h] 
	\centering
	\includegraphics[width=0.5 \textwidth]{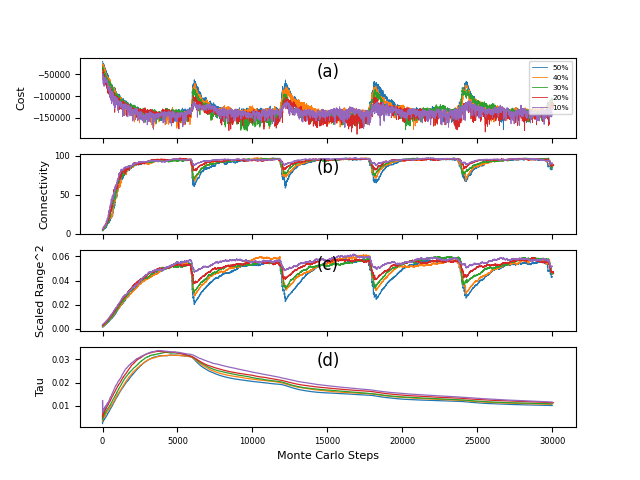}
	\caption{Variation diagrams for (a) cost (b) connectivity (c) average scaled ranges$^2$ (d) average Tau over Monte Carlo steps is demonstrated for a mobile ad hoc network facing node failures. The network is created with the node distribution density $\rho = 0.05$. The diagram's colors represent the different node percentages of failed nodes. These diagrams represent the average of ten statistical ensembles of the intended network.}
	\label{fig:movement_compare_four_failure}
\end{figure}

\begin{figure}[!h] 
	\centering
	\includegraphics[width=0.5 \textwidth]{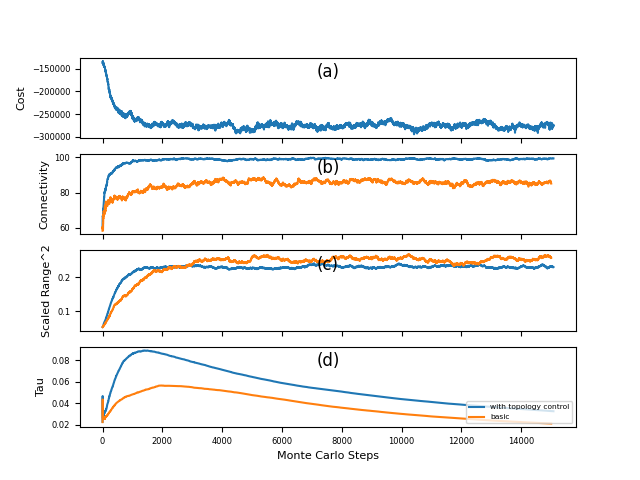}
	\caption{Variation diagrams for (a) cost (b) connectivity (c) average scaled ranges$^2$ (d) average Tau over Monte Carlo steps is demonstrated for a 3D mobile ad hoc network with the node distribution density $\rho = 0.05$. These diagrams represent the average of ten statistical ensembles of the intended network.}
	\label{fig:movement_compare_four_3d}
\end{figure}

\section{Conclusion and Future Works}
\label{sec:conclusion}
In light of the growing number of IoT applications and the necessity of managing the topology of these networks to enhance their efficiency, we introduce a novel self-organizing approach for topology control of ad hoc-based IoT networks. 
This method is proposed by considering the network under investigation as a complex physics system and defining a Hamiltonian function for this system. In this study, the Hamiltonian function is formulated in a way that will cause the network dynamics to develop in a manner that will change the topology of the network from its initial form to the desired topology.
We have evaluated our proposed method for cost, connectivity, and energy usage through three topology types with various numbers of parameters.
The results obtained from employing this method demonstrate that it can achieve the network with the near $100 \% $ with the reduced amount of energy consumption while preserving it in that state so that minor local changes through the network don't affect the connectivity. 

In addition, this approach attempts to improve the network's link stability and even promises that, in case if a significant number of the network's nodes go away or failed the network will promptly and self-organizingly restore optimal connectivity. Thus, there can always be effective communication throughout the network. 
By enhancing connectivity, time and loss during data transfers are reduced, and the reliability of the network is improved. 
The proposed model can be extended to 3-dimensions, which include a wide range of exciting IoT applications.

A number of benefits are associated with the proposed method. It is adaptive to environments with different device densities, levels of mobility, and energy source availability. It self-organizes and reaches a stable state without the help of outside assistance. It is energy-efficient due to topological modifications, so it can be integrated with various routing protocols and specific communication models. The model can be applied to a wide range of devices without requiring any changes on the hardware level. It is robust to local and global changes and capable of compensating for it. For ad hoc networks, it is a fully distributed method. 
These characteristics work together to make the network scalable, indicating that adding or removing devices from the network is straightforward because it quickly adapts and recovers its robust connectivity. 
Also, the proposed model is not a application-specific network model; rather, it is a general model with only a limited set of control parameters.

Future research may deal with: (1) Modifying the Hamiltonian function for specific IoT case studies and increase problem solving accuracy by adding any physical practical constraints, like interference, delay or communication limitations, related to the application case. (2) Adding real movements to the nodes, employing urban mobility generators. (3) Employing machine learning techniques to learn the appropriate $\gamma$ for all possible values of $T$ and $\rho$.

\bibliography{bibliography}

\end{document}